\documentclass[10pt,conference]{IEEEtran}

\usepackage[hidelinks]{hyperref}

\usepackage{makecell}
\usepackage{fixltx2e}
\usepackage{graphicx}
\usepackage{caption}
\usepackage{subcaption}

\IEEEoverridecommandlockouts
\usepackage{amssymb,amsmath,amsthm}
\usepackage[font=small,labelfont=bf]{caption}

\usepackage{comment}

\makeatletter
\newcommand{\biggg}[1]{{\hbox{$\left#1\vbox to 20.5pt{}\right.\n@space$}}}
\newcommand{\Biggg}[1]{{\hbox{$\left#1\vbox to 23.5pt{}\right.\n@space$}}}
\newcommand{\bigggg}[1]{{\hbox{$\left#1\vbox to 26.5pt{}\right.\n@space$}}}
\newcommand{\Bigggg}[1]{{\hbox{$\left#1\vbox to 29.5pt{}\right.\n@space$}}}
\newcommand{\biggggg}[1]{{\hbox{$\left#1\vbox to 32.5pt{}\right.\n@space$}}}
\newcommand{\Biggggg}[1]{{\hbox{$\left#1\vbox to 35.5pt{}\right.\n@space$}}}
\newcommand{\bigggggg}[1]{{\hbox{$\left#1\vbox to 38.5pt{}\right.\n@space$}}}
\newcommand{\Bigggggg}[1]{{\hbox{$\left#1\vbox to 41.5pt{}\right.\n@space$}}}
\makeatother

\usepackage{bm,cite,algorithm,algorithmic,float,amsmath,amssymb}
\usepackage{mdwmath}
\usepackage{mdwtab}
\usepackage{xcolor}
\usepackage{cool}
\usepackage{makeidx,enumitem}
\usepackage{balance}
\floatname{algorithm}{Algorithm}

\makeatletter

\usepackage{multirow}
\usepackage{epstopdf}
\makeatletter
\newcommand\notsotiny{\@setfontsize\notsotiny{7}{6.5}}
\makeatother
\begin{document}
\title{\vspace{-1mm} An Ultra-Wideband Study of Vegetation Impact on Upper Midband / FR3 Communication\vspace{-3mm}}

\author{
    \IEEEauthorblockN{Naveed A. Abbasi, Tathagat Pal, Kelvin Arana, Vikram Vasudevan, Jorge Gomez-Ponce, \\ Young Han Nam, Charlie Zhang and Andreas F. Molisch}
    \thanks{N. A. Abbasi, T. Pal, K. Arana, V. Vasudevan, J. Gomez-Ponce and A. F. Molisch are with University of Southern California, Los Angeles, CA, USA. 
	Email: \{nabbasi, tpal, aranaoro, vikramva, gomezpon, molisch\}@usc.edu. J. Gomez-Ponce is also with ESPOL Polytechnic University, Guayaquil, Ecuador. Young Han Nam and Charlie Zhang are with Samsung Research America, Richardson, TX, USA.}
}

 \vspace{-100mm}

\maketitle
\vspace{-90mm}
\begin{abstract}
Growing demand for high data rates is driving interest in the upper mid-band (FR 3) spectrum (6-24 GHz). While some propagation measurements exist in literature, the impact of vegetation on link performance remains under-explored. This study examines vegetation-induced losses in an urban scenario across 6-18 GHz. A simple method for calculating vegetation depth is introduced, along with a model that quantifies additional attenuation based on vegetation depth and frequency, divided into 1 GHz sub-bands. We see that excess vegetation loss increases with vegetation depth and higher frequencies. These findings provide insights for designing reliable, foliage-aware communication networks in FR 3.
\end{abstract}
\section{Introduction}
The rising demand for high data rates in urban areas drives interest in the upper mid-band spectrum (6-24 GHz), also called Frequency Range 3 (FR3) by the cellular standardization organization 3GPP \cite{kang2024cellular}. Accurate channel characterization through precision measurements is critical for effective system design. However, given uncertainty about which bands will be allocated for high-data-rate use, ultra-wideband (UWB)\footnote{UWB is defined here as having over 20\% relative bandwidth, consistent with one FCC definition. Although the FCC also classifies transmissions with more than 500 MHz bandwidth as UWB, this is based on regulatory factors rather than channel statistics \cite{molisch2009ultra}.} measurements across the whole band are vital to ensure comprehensive, flexible data collection. Additionally, understanding the frequency-dependence of channel model parameters for the 3GPP- and other channel models is essential for realistic system performance assessment \cite{3gppFR3}.

There have been a number of propagation measurements in the upper mid-band.\footnote{Only sample papers are discussed due to space constraints.} These include cases such as measurement of specific propagation effects, like transmission and reflection from building materials \cite{shakya2024wideband}; single-link-end measurements, often using omni-directional antennas, which can be narrowband \cite{oh2019empirical}; directional measurements at one link end \cite{miao2023sub}; and double-directional measurements, which are typically wideband \cite{shakya2024urban,abbasi2025ultrawideband}.

Even though upper-midband systems will be deployed mainly in densely populated areas, analysis of the impact of vegetation is important, as many suburban, urban, and even metropolitan areas contain significant vegetation, such as tree-lined avenues and plazas with bushes and trees. Such vegetation introduces significant losses due to attenuation and scattering, which are expected to worsen with higher frequencies and greater foliage density. Despite this, measurement-based studies on the impact of vegetation on communication links in the upper midband remain rare. To the best of our knowledge, the only in-depth study is \cite{schwering1988millimeter}, which however only measured narrowband at 9.6 GHz and compared the results to 29 and 59 GHz results. While \cite{ITUR833} proposes vegetation attenuation models for frequencies from 30 MHz to 100 GHz, they are only based on measurements below 3 GHZ and above 30 GHz, while models in between are based on theoretical considerations only.

To address this gap, the current paper employs UWB measurements across the 6-18 GHz range to investigate the excess losses caused by varying vegetation depths in urban environments for the first time in the literature. We propose a simple method to calculate vegetation depth and model its effects on communication links, analyzing data in one-GHz-wide sub-bands. By modeling these additional vegetation-induced losses with a linear fit, we develop a practical vegetation loss model that quantifies the excess attenuation as a function of vegetation depth and frequency, providing crucial insights for the design and optimization of future mid-band communication systems.

\section{Measurement Setup and Site}
\subsection{Testbed description}
For the current measurement campaign, we use a custom RF-over-fiber (RFoF) frequency-domain channel sounder, as detailed in \cite{abbasi2025ultrawideband}. The sounder is based on a vector network analyzer (VNA) covering a 12 GHz bandwidth from 6 to 18 GHz, with high-gain HGHA618 horn antennas (6-18 GHz) mounted on high-precision rotors. The transmitter (Tx) is mounted on a building at a height of over 20 meters, emulating a base station (BS), while the receiver (Rx), simulating user equipment (UE), is positioned 2 meters above the ground (to ensure a travel through the canopy of the tree). The Tx remained stationary throughout the measurements. Each VNA sweep records 12,001 frequency points across a 12 GHz bandwidth, enabling the measurement of excess delays up to 1 $\mu$s. This corresponds to a maximum multipath propagation distance of 300 meters, sufficient to avoid aliasing in a microcellular scenario in an urban environment, which is the focus of this study.

The setup aligns the Tx and Rx horn antennas to point to each other along the line-of-sight (LoS), establishing a coordinate system in which 
 the LoS multipath component (MPC) has a co-elevation of $0^\circ$ at Tx and Rx. The Rx was rotated in the azimuthal range of $-60^\circ$ to $60^\circ$ in $10^\circ$ steps and in the elevation range of $-30^\circ$ to $30^\circ$ in $10^\circ$ steps to identify potential errors from misalignment. However, no significant alignment errors were observed. To isolate ``system and antenna" effects from the measurements, a time-gated over-the-air (OTA) calibration was performed daily throughout the campaign at a calibration point located 44 meters from the Tx, as illustrated in Fig. \ref{fig:site} while ensuring there is no vegetation within the first Fresnel zone from the Tx to this calibration location. Additionally, the measurements were conducted at night in a cordoned-off area to ensure a static and controlled environment throughout the measurement process.
\subsection{Measurement site}
\begin{figure}[t!]
	\centering
 \vspace{1mm}
	\includegraphics[width=0.7\linewidth]{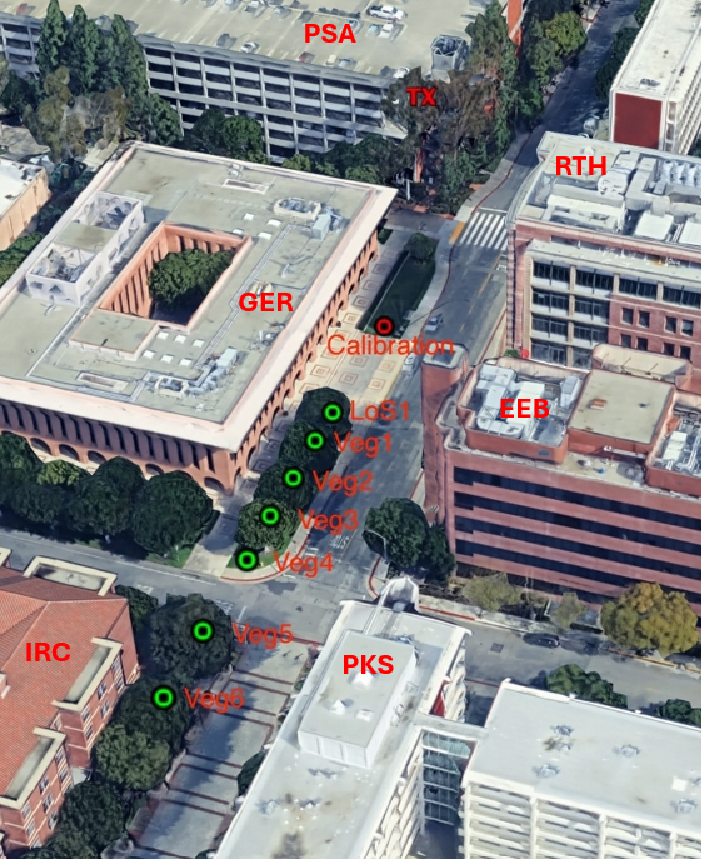}
	\caption{Measurement site.}
	\label{fig:site}
\end{figure}
The measurements were conducted on the University of Southern California's University Park Campus in Los Angeles, CA, USA, specifically along McClintock Avenue, as shown in Fig. \ref{fig:site}. The Tx was positioned near the top of the PSA building on an external staircase, providing an elevated vantage point for LoS signal transmission. The Rx positions, labeled as $Veg1$ through $Veg6$, were strategically placed along McClintock Avenue and nearby areas to study the impact of vegetation and urban obstructions on the communication channel. The calibration point was located directly across from the Tx on the opposite side of McClintock Avenue.

The environment represents an urban street canyon scenario, featuring building facades (e.g., GER, EEB, IRC, PKS), foliage of varying densities, and parked vehicles, making it ideal for studying vegetation-induced losses. Among the Rx positions, $LoS1$ (\(64.5 \, \mathrm{m}\) from the Tx) represents a clear LoS path with no vegetation obstruction. In contrast, $Veg1$ (\(74.7 \, \mathrm{m}\)) through $Veg6$ (\(126 \, \mathrm{m}\)) introduced varying vegetation depths. It is important to note that longer distances do not always correspond to greater vegetation depth, as the vegetation a signal traverses depends on the specific scenario geometry. For instance, the LoS to some farther Rx points may go over the canopies of many of the trees and just correspond to lower vegetation depth.   

The vegetation depth for each Rx point, and the corresponding excess losses due to vegetation, are described in the next section. A summary of distances between the Tx and Rx points is provided in Table \ref{table:rx_points}.
\begin{table}[b]
\caption{Rx point description.}
\begin{tabular}{|c|c|c|}
\hline
\textbf{Identifier} & \textbf{Tx-Rx Distance (m)} & \textbf{Vegetation Depth (m)} \\ \hline \hline
$LoS1$              & 64.5                        & 0                             \\ 
$Veg1$              & 74.7                        & 5.18                          \\ 
$Veg2$              & 83.1                        & 15.59                         \\ 
$Veg3$              & 90.9                        & 22.18                         \\ 
$Veg4$              & 97.5                        & 27.82                         \\
$Veg5$              & 116.7                       & 26.85                         \\ 
$Veg6$              & 126                         & 24.31                         \\ \hline
\end{tabular}
\label{table:rx_points}
\end{table}
\subsection{Calculation of Vegetation Depth}
\begin{figure}[t!]
    \centering
    \includegraphics[width=0.9\linewidth]{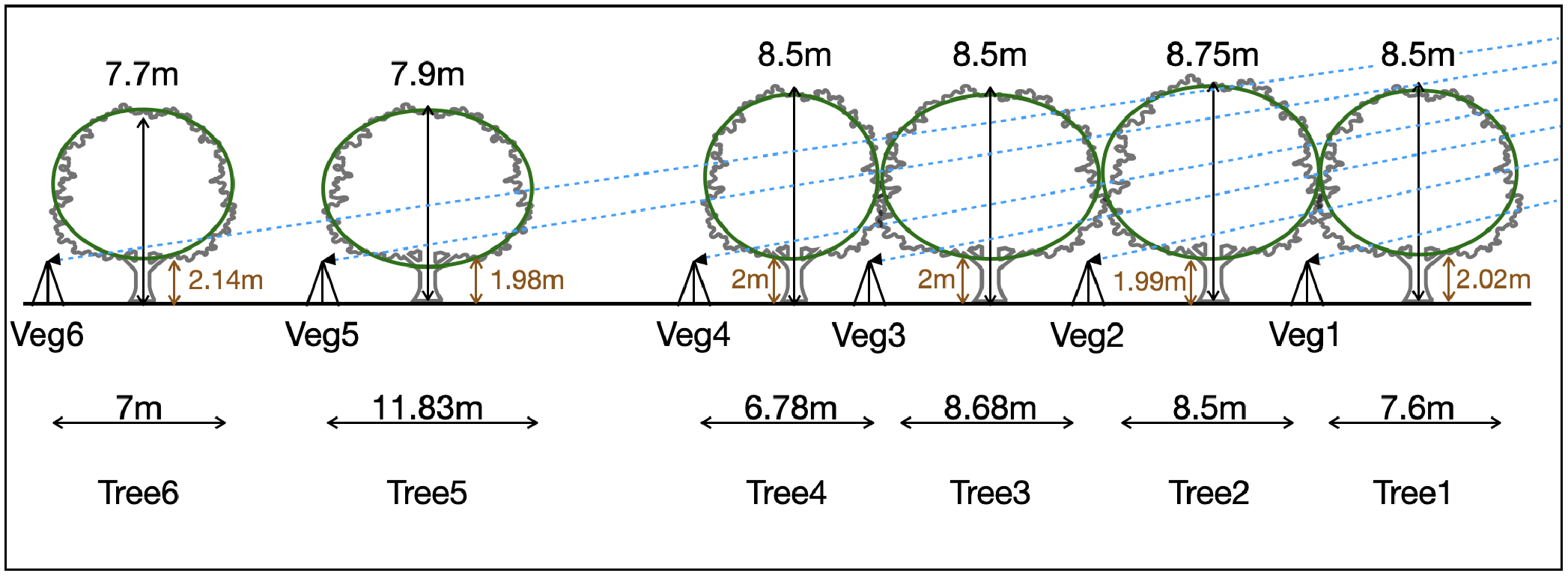}
    \caption{Elliptical modeling of tree silhouettes for vegetation depth calculation.}
    \label{fig:depth_model_ellipse}
\end{figure}
The geometry of each tree was measured on-site using rods and GoPro cameras, allowing for precise determination of tree dimensions, including trunk height, foliage height, and tree width. These measurements were used to model the silhouette of each tree. An ellipse was chosen to represent the tree's silhouette as it provides a more realistic approximation compared to a circle. Modeling a tree as a circle can lead to inaccuracies if the width and height of the canopy are not similar: if the circle is inscribed within the tree boundary, it underestimates the foliage depth by ignoring the outer edges; conversely, if the circle is expanded to cover the boundary, it overestimates the depth by extending beyond the tree's actual shape. Additionally, modeling the intricate structure of branches and finer foliage is not practical. Therefore, the elliptical model offers a balanced and effective representation of the tree's main foliage, as illustrated in Fig. \ref{fig:depth_model_ellipse}.

To calculate the vegetation depth, we establish the equations of both the LoS line and the tree ellipse, and solve for the points of intersection. From this, the distance between them is easily determined. This process is repeated for each tree along the signal path, for all Rx locations, to compute the total vegetation depth. For some points, such as $Veg6$, the signal path passes through multiple trees, requiring the calculation of vegetation depth as the sum of distances through each ellipse.







The results are summarized in Table \ref{table:rx_points}.

\section{Parameters and Processing}
The measurement setup generates frequency-domain scans across various Tx-Rx configurations, resulting in a three-dimensional tensor \( H_{meas}(f, \phi_{Rx}, \tilde{\theta}_{Rx}; d) \), where \( f \) represents the frequency, \( \phi_{Rx} \) denotes the Rx azimuth orientations, \( \tilde{\theta}_{Rx} \) is the Rx elevation angle, and \( d \) is the Tx-Rx distance. To obtain the calibrated channel transfer function, the measured channel response is normalized by the OTA calibration, \( H_{OTA}(f) \), which removes the effect of the measurement system, antennas (gain only while directional filtering remains), and the impact of measuring at OTA distance \( d_{OTA} \) as:
\begin{align}
    H(f, \phi_{Rx}, \tilde{\theta}_{Rx}; d) &= 
    \frac{H_{meas}(f, \phi_{Rx}, \tilde{\theta}_{Rx}; d)}
    {H_{OTA}(f)}.
\end{align}

Directional power delay profiles (PDP) are computed using the inverse Fourier transform of the calibrated channel transfer function:
\begin{multline}
    P_{calc}(\tau, \phi_{Rx}, \tilde{\theta}_{Rx}; d) = 
    \left| \mathcal{F}_{f}^{-1} \left\{ H(f,\phi_{Rx}, \tilde{\theta}_{Rx}; d) \right\} 
    \right|^2.    
\end{multline}

Noise reduction is performed using thresholding and delay gating techniques, as detailed in \cite{gomez2023impact}, and is defined as:
\begin{align}
    P(\tau) &= \left[ P_{calc}(\tau) : (\tau \leq \tau_{gate}) \land 
    (P_{calc}(\tau) \geq P_{\lambda}) \right],
\end{align}
where \( \tau_{gate} \) is the maximum delay gate, and \( P_{\lambda} \) is the power threshold set to remove noise (set at \(12~\mathrm{dB}\) above noise floor).

Since we aim to assess the impact of vegetation, it is crucial to isolate the direct path between the Tx and Rx that avoids interactions with other objects in the scenario. To achieve this, we focus on the direct LoS path between the Tx and Rx, analyzing the directional PDPs to determine the optimal alignment. This alignment is guided by the scenario's geometry and matched with the delay domain expectation of the direct LoS path, enabling us to accurately identify the corresponding directional PDP component. The power of this component, corresponding to the delay \(\tau_{LoS} = d / c\) (where \(c\) is the speed of light), is denoted as \(P_{LoS}(f; d)\).

While the high delay resolution might suggest that isolation of the strongest path could be achieved using either a constructed omni-directional or best-aligned PDP, further investigations revealed that the omni-directional PDPs contained additional contributions even in the delay bin of the LoS path that remain unresolvable at the Fourier delay grid resolution. These contributions bypass the vegetation and thus distort the results. By contrast, analysis of the best-aligned PDP indicated an absence of significant alignment errors, as mentioned above. For this reason, all further processing is based on the best-aligned PDP and, in particular, the LoS path component extracted from it. We note, however, that despite the use of directional PDPs and high delay resolution, residual small-scale fading caused by interference from unresolvable MPCs is present (and inevitable), leading to fluctuations in the observed results. Such MPCs might arise from (unresolvable) scattering, diffraction around the canopy, and ground reflections of MPCs passing through the canopy; they will arise not only in our measurements but in all real-world urban scenarios with vegetation.  

The excess vegetation loss is then calculated as the difference between \(P_{LoS}(f; d)\), and the theoretical Friis received power in dB (\(P_{Friis}(f, d)\)) at the same distance (\(d\)) and frequency, as described in \cite[Chapter 4]{molisch2023wireless}, and is expressed as:
\begin{equation}
L_{veg}(f; d) = P_{Friis}(f, d) - P_{LoS}(f; d).
\end{equation}

To model the relationship between excess vegetation loss and vegetation depth, we fit a linear model as:
\begin{equation}
L_{veg}(f, d_{veg}) = \alpha (f) d_{veg},
\end{equation}
where \(d_{veg}\) is the vegetation depth i.e. the portion of the total distance \(d\) that is obstructed by vegetation (calculated values for the current measurements in Table \ref{table:rx_points}), and \(\alpha(f)\) is the slope of the fitted line that corresponds to excess vegetation loss per unit distance. Please note that the linear fit is constrained to start from the origin \((d_{veg} = 0, L_{veg} = 0)\), as no additional loss due to vegetation should be observed when vegetation depth is zero, corresponding to scenarios without vegetation in the channel (e.g., $LoS1$).  

\section{Results}
\subsection{Power delay profiles}
\begin{figure}[t!]
\centering
    \begin{subfigure}[b]{0.72\columnwidth}
    \centering
        \includegraphics[width=1\columnwidth]{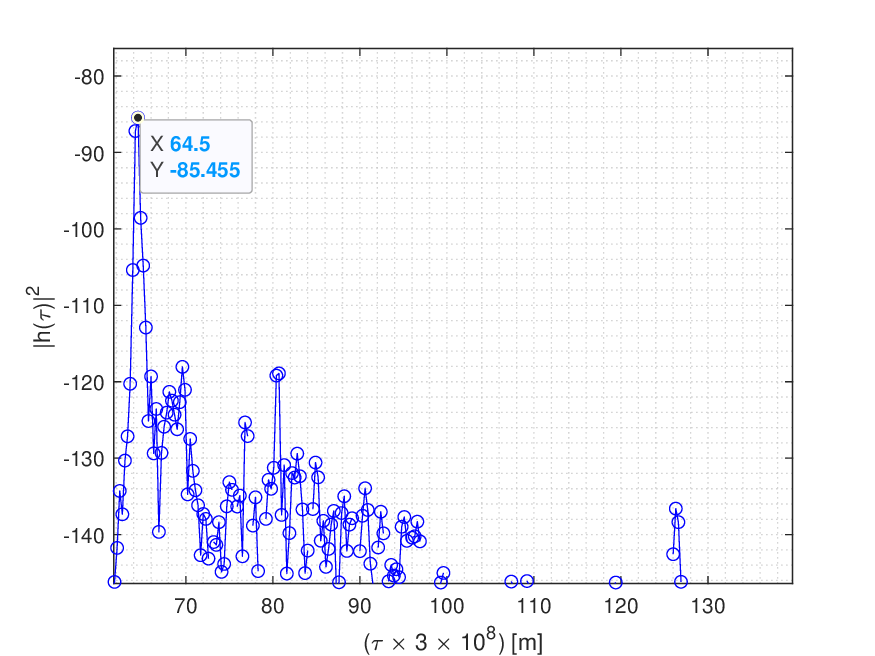}
        \caption{PDP for $LoS1$.}
        \vspace*{0mm}
        \label{fig:pdp-los}
        \end{subfigure}
    \begin{subfigure}[b]{0.72\columnwidth}
    \centering
        \includegraphics[width=1\columnwidth]{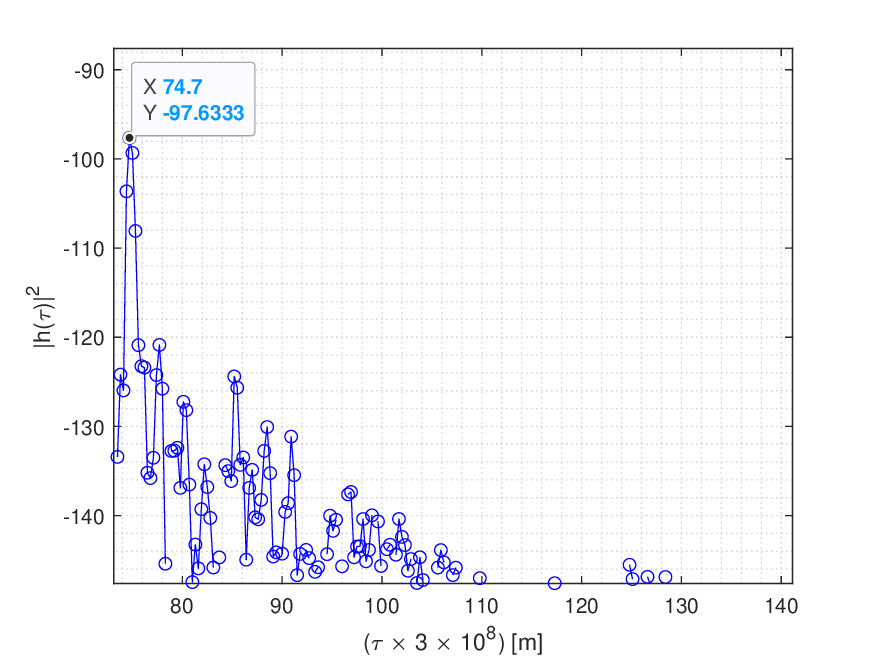}
        \caption{PDP for $Veg1$.}
        \vspace*{0mm}
        \label{fig:pdp-veg1}
        \end{subfigure}
\caption{Sample PDPs between 6-7 GHz.}%
\label{fig:pdp}%
\vspace{-0 mm}
\vspace{-5mm}
\end{figure}
Two sample PDPs for 6–7 GHz are shown in Fig. \ref{fig:pdp-los} and Fig. \ref{fig:pdp-veg1}. The delay on the x-axis is expressed in meters, calculated as \(\tau \times c\), where \(c = 3 \times 10^8 \, \mathrm{m/s}\), providing a clear representation of the propagation paths.

In both cases, the strongest component, corresponding to the LoS, appears at a delay matching the physical Tx-Rx distance, verifying the geometry of the scenario. Additional components, representing reflections within the environment, are also visible but are not the focus here. Focusing on the primary LoS component, we see in Fig. \ref{fig:pdp-los} ($LoS1$, no vegetation) that the power of this component is \(-85.46 \, \mathrm{dB}\), close to the expected Friis received power of \(-84.89 \, \mathrm{dB}\) at 6.5 GHz. In contrast, Fig. \ref{fig:pdp-veg1} ($Veg1$, \(d_{veg} \approx 5 \, \mathrm{m}\)) shows the LoS power at \(-97.63 \, \mathrm{dB}\), compared to the Friis received power of \(-86.16 \, \mathrm{dB}\). This indicates an excess loss of nearly \(11.5 \, \mathrm{dB}\) due to vegetation.
\subsection{Vegetation depth vs excess loss}
\begin{figure}[t!]
	\centering
 \vspace{1mm}
	\includegraphics[width=0.72\linewidth]{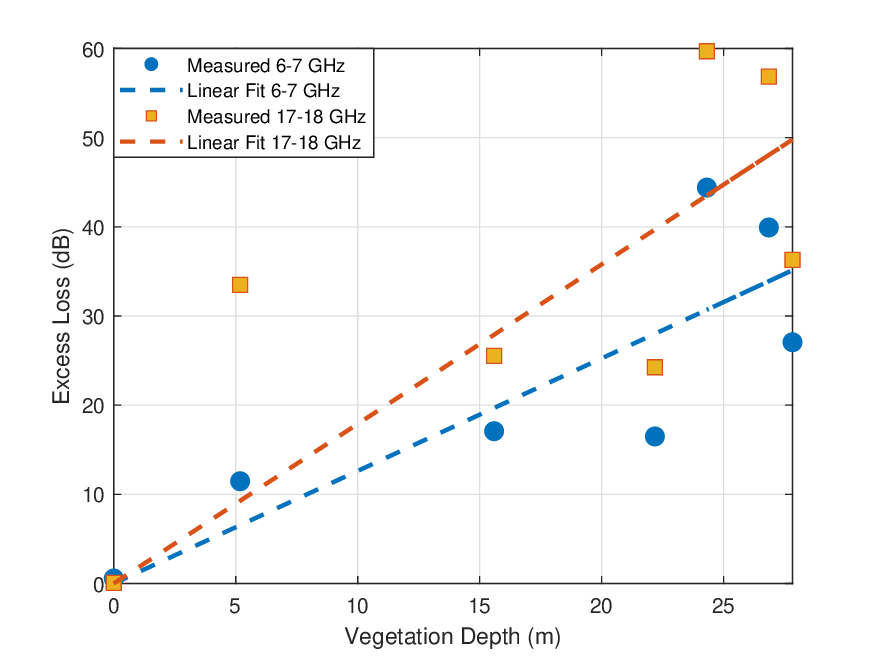}
	\caption{Vegetation depth vs excess loss for different frequencies.}
	\label{fig:veg_depth_vs_loss}
\end{figure}
\begin{table}[h!]
\centering
\caption{Linear fitting for excess loss vs vegetation depth.}
\label{tab:excess_loss_fit}
\begin{tabular}{|c||c|c|c|}
\hline 
\textbf{Frequency} & $\alpha_{min}$ & $\alpha$ & $\alpha_{max}$ \\ \hline \hline
6-7 GHz & 0.86 & 1.26 & 1.66 \\ 
7-8 GHz & 1.04 & 1.50 & 1.95 \\ 
8-9 GHz & 1.11 & 1.55 & 2.00 \\ 
9-10 GHz & 0.91 & 1.57 & 2.23 \\ 
10-11 GHz & 1.24 & 1.68 & 2.11 \\ 
11-12 GHz & 1.23 & 1.80 & 2.38 \\ 
12-13 GHz & 1.33 & 1.80 & 2.26 \\ 
13-14 GHz & 1.12 & 1.69 & 2.27 \\ 
14-15 GHz & 1.35 & 1.80 & 2.25 \\ 
15-16 GHz & 1.22 & 1.81 & 2.40 \\ 
16-17 GHz & 1.25 & 1.76 & 2.26 \\ 
17-18 GHz & 1.10 & 1.79 & 2.48 \\ 
\hline
\end{tabular}
\end{table}
Fig. \ref{fig:veg_depth_vs_loss} shows the fitting results for the 6-7 GHz and 17-18 GHz bands where we see that excess loss generally increases with vegetation depth, and \(\alpha\) varies with frequency. Moreover, we see that at both frequencies, for a vegetation depth of 0, the excess loss is also nearly zero, consistent with our model expressed in equations (4) and (5). While the linear fits for these bands capture the overall trend, the measured data points exhibit noticeable scattering around the fitted line. This variability suggests that additional measurements may be needed to refine the model further. However, as an initial model, the results effectively demonstrate the increasing loss with vegetation depth and highlight the frequency-dependent nature of attenuation, with higher frequencies (e.g., 17-18 GHz) experiencing more loss. Table \ref{tab:excess_loss_fit} summarizes the slopes \(\alpha\) for different frequency bands. The values of \(\alpha\) range from approximately 1.26 \(\mathrm{dB/m}\) at 6–7 GHz to 1.79 \(\mathrm{dB/m}\) at 17–18 GHz, reflecting the frequency-dependent nature of foliage-induced attenuation. The variation of \(\alpha\) within each band, represented by \(\alpha_{min}\) and \(\alpha_{max}\), accounts for the inherent small-scale fading observed during measurements as well as the limitations of the Fourier delay resolution.
\section{Conclusion}
This study is the first ultra-wideband investigation of vegetation-induced losses in the upper mid-band spectrum (6–18 GHz) under urban conditions. We see a general trend of increasing excess loss with vegetation depth, with higher frequencies experiencing more attenuation. While the proposed model effectively captures this behavior, noticeable variability in the measurements suggests that additional measurements are necessary to further validate and refine the results based on scenario, the vegetation type, foliage density etc.

\begin{thebibliography}{10}
\providecommand{\url}[1]{#1}
\csname url@samestyle\endcsname
\providecommand{\newblock}{\relax}
\providecommand{\bibinfo}[2]{#2}
\providecommand{\BIBentrySTDinterwordspacing}{\spaceskip=0pt\relax}
\providecommand{\BIBentryALTinterwordstretchfactor}{4}
\providecommand{\BIBentryALTinterwordspacing}{\spaceskip=\fontdimen2\font plus
\BIBentryALTinterwordstretchfactor\fontdimen3\font minus \fontdimen4\font\relax}
\providecommand{\BIBforeignlanguage}[2]{{%
\expandafter\ifx\csname l@#1\endcsname\relax
\typeout{** WARNING: IEEEtran.bst: No hyphenation pattern has been}%
\typeout{** loaded for the language `#1'. Using the pattern for}%
\typeout{** the default language instead.}%
\else
\language=\csname l@#1\endcsname
\fi
#2}}
\providecommand{\BIBdecl}{\relax}
\BIBdecl

\bibitem{kang2024cellular}
S.~Kang, M.~Mezzavilla, S.~Rangan, A.~Madanayake, S.~B. Venkatakrishnan, G.~Hellbourg, M.~Ghosh, H.~Rahmani, and A.~Dhananjay, ``Cellular wireless networks in the upper mid-band,'' \emph{IEEE Open Journal of the Communications Society}, 2024.

\bibitem{molisch2009ultra}
A.~F. Molisch, ``Ultra-wide-band propagation channels,'' \emph{Proceedings of the IEEE}, vol.~97, no.~2, pp. 353--371, 2009.

\bibitem{3gppFR3}
{RAN 1}, ``Rel-19 ran1 agreements - 7-24 ghz channel modeling,'' 3GPP, Tech. Rep., 2025.

\bibitem{shakya2024wideband}
D.~Shakya, M.~Ying, T.~S. Rappaport, H.~Poddar, P.~Ma, Y.~Wang, and I.~Al-Wazani, ``Wideband penetration loss through building materials and partitions at 6.75 ghz in fr1 (c) and 16.95 ghz in the fr3 upper mid-band spectrum,'' \emph{arXiv preprint arXiv:2405.01362}, 2024.

\bibitem{oh2019empirical}
S.-S. Oh, J.-W. Choi, H.-C. Lee, Y.-C. Lee, B.-L. Cho, I.-Y. Lee, J.-H. Lim, J.-I. Lee, and S.~W. Park, ``An empirical propagation prediction model for urban street canyon environments at 6, 10, and 18 ghz,'' \emph{Microwave and Optical Technology Letters}, vol.~61, no.~6, pp. 1574--1578, 2019.

\bibitem{miao2023sub}
H.~Miao, J.~Zhang, P.~Tang, L.~Tian, X.~Zhao, B.~Guo, and G.~Liu, ``Sub-6 ghz to mmwave for 5g-advanced and beyond: Channel measurements, characteristics and impact on system performance,'' \emph{IEEE Journal on Selected Areas in Communications}, vol.~41, no.~6, pp. 1945--1960, 2023.

\bibitem{shakya2024urban}
D.~Shakya, M.~Ying, T.~S. Rappaport, P.~Ma, I.~Al-Wazani, Y.~Wu, Y.~Wang, D.~Calin, H.~Poddar, A.~Bazzi \emph{et~al.}, ``Urban outdoor propagation measurements and channel models at 6.75 ghz fr1 (c) and 16.95 ghz fr3 upper mid-band spectrum for 5g and 6g,'' \emph{arXiv preprint arXiv:2410.17539}, 2024.

\bibitem{abbasi2025ultrawideband}
\BIBentryALTinterwordspacing
N.~A. Abbasi, K.~Arana, J.~Gomez-Ponce, T.~Pal, V.~Vasudevan, A.~Bist, O.~G. Serbetci, Y.~H. Nam, C.~Zhang, and A.~F. Molisch, ``Ultra-wideband double-directionally resolved channel measurements of line-of-sight microcellular scenarios in the upper mid-band,'' 2024. [Online]. Available: \url{https://arxiv.org/abs/2412.12306}
\BIBentrySTDinterwordspacing

\bibitem{schwering1988millimeter}
F.~K. Schwering, E.~J. Violette, and R.~H. Espeland, ``Millimeter-wave propagation in vegetation: Experiments and theory,'' \emph{IEEE Transactions on Geoscience and Remote Sensing}, vol.~26, no.~3, pp. 355--367, 1988.

\bibitem{ITUR833}
I.~T. Union, ``Recommendation {ITU-R P.833-10} attenuation in vegetation,'' ITU, Tech. Rep., 2021.

\bibitem{gomez2023impact}
J.~Gomez-Ponce, N.~A. Abbasi, Z.~Cheng, S.~Abu-Surra, G.~Xu, J.~Zhang, and A.~F. Molisch, ``Impact of noisy measurements with fourier-based evaluation on condensed channel parameters,'' \emph{IEEE Transactions on Wireless Communications}, 2023.

\bibitem{molisch2023wireless}
A.~F. Molisch, \emph{Wireless communications}, 3rd~ed.\hskip 1em plus 0.5em minus 0.4em\relax IEEE Press - John Wiley \& Sons, 2023.

\end{thebibliography}

\end{document}